\documentstyle[aps,twocolumn,prl,epsf]{revtex}

\renewcommand{\ref}[1]{\raisebox{.6ex}{[#1]}}

\newcommand{\be}{\begin{equation}}
\newcommand{\ee}{\end{equation}}

\newcommand{\bea}{\begin{eqnarray}}
\newcommand{\eea}{\end{eqnarray}}

\newcommand{\ba}{\begin{array}}
\newcommand{\ea}{\end{array}}

\begin{document}


\twocolumn[
\hsize\textwidth\columnwidth\hsize\csname @twocolumnfalse\endcsname

\title{ Quantum Spinodal Decomposition in 
        Multicomponent Bose-Einstein Condensates }

\author{   S.T. Chui    \\
          Bartol Research Institute \\
         University of Delaware, Newark, DE 19716, USA \\
                 and \\
             P. Ao \\
         Department of Theoretical Physics  \\
         Ume\aa{\ }University, 901 87 Ume\aa,  Sweden        }

\maketitle

\widetext

\begin{abstract}
We investigate analytically
the non-equilibrium spatial phase segregation process of a mixture of 
alkali Bose-Einstein condensates. Two stages (I and II)
are found in analogy to the classical spinodal decomposition.
The coupled non-linear Schr\"odinger equations enable us to give
a quantitative picture of the present dynamical process in a square well trap. 
Both time and length scales in the stage I are obtained.
We further propose that the stage II is dominated by the Josephson
effect between different domains of same condensate
different from scenarios in the classical spinodal decomposition.
Its time scale is estimated. 

\noindent
PACS${\#}$:  03.75.Fi; 64.75.+g 
 
\end{abstract}
] 

 
\narrowtext

The recent realizations of two \cite{jila1,jila2} and three \cite{mit}
component alkali Bose-Einstein condensates (BEC's) in one trap
provide us with new systems to explore the physics in otherwise 
unachievable parameter regimes \cite{law,ho,cta}.
These systems have two noticeable advantages:
the easy control of experimental parameters and the relative simplicity 
of the mathematical description.
A direct comparison between theoretical calculations and
experimental observations can be made.
The purpose of the present paper is to explore 
one of main questions in the non-equilibrium statistical physics using those
new systems: spinodal decomposition in a binary-solution system.
It is a typical example of phase ordering dynamics:
the growth of order through domain coarsening 
when a system is quenched from the homogeneous phase into 
a broken-symmetry phase \cite{cahn}.
Systems quenched from a disordered phase into
an ordered phase do not order instantaneously.
Instead, different length scales set in as the domains form and
grows with time, 
and different broken symmetry phases compete to select
the equilibrium state.
We show that 
it is possible to have an analogous spinodal decomposition in BEC's, 
which manifests all the main phenomenology except those 
constrained by the trap size.
In the following, 
we shall study the main features in this dynamical evolution process, 
starting from the homogeneous unstable state of BEC's.
To differentiate the present situation from the usual one, we shall call the 
present one the quantum spinodal decomposition, and the previous ones
the classical spinodal decomposition.

We start from the time dependent
non-linear Schr\"odinger equations
\bea
    i \hbar \frac{\partial }{\partial t}  \psi_1 & = &  
       \left[ - \frac{\hbar^2}{2m_1} \nabla^2  
       + (U_1(x) - \mu_1 )
       + G_{11} |\psi_1|^2 \right] \psi_1 \nonumber  \\
    & & + G_{12} |\psi_2|^2 \psi_1 \; ,
\eea
and 
\bea
    i \hbar \frac{\partial }{\partial t} \psi_2 & = & 
      \left[- \frac{\hbar^2}{2m_2} \nabla^2 
      + (U_2(x) - \mu_2 )
      + G_{22} |\psi_2|^2 \right] \psi_2  \nonumber \\ 
    & & + G_{12} |\psi_1|^2 \psi_2  \; .
\eea
Here $\psi_j(x,t)$,  $m_j$,  $U_j$ with $j=1,2$ are
the effective wave function,  the mass, 
and the trapping potential of the $j$th condensate.
The repulsive 
interaction between the $j$th condensate atoms is specified by $G_{jj}$,
and that between 1 and 2  by $G_{12}$.
The Lagrangian multipliers, the chemical potentials $\mu_1$ and $\mu_2$, 
are fixed by the relations
$
   \int d^3 x  |\psi_j(x,t)|^2  =  N_j \; , j =1,2
$,
with $N_j$ the number of the $j$th condensate atoms.
Eq. (1) and (2) are mean field equations, since we treat the effective 
wave functions as $c$ numbers, corresponding to
the Hartree-Fock-Bogoliubov approximation.
They provide a good description for the slow dynamics in both
alkali BEC's and the superfluid He4 \cite{nlse} on length
scale larger than the range of microscopic interaction. 

Experimentally, the trapping potentials $\{ U_j \}$ are
simple harmonic in nature.
For the sake of simplicity and to illustrate the physics
we shall consider a square well trapping potential $U_j = U$:
zero inside and large (infinite) outside,
unless otherwise explicitly specified.

We consider the strong mutual repulsive interaction
\be
   G_{12} > \sqrt{ G_{11} G_{22} }   \; .
\ee
In this regime the equilibrium state for two Bose-Einstein condensates
is a spatial segregation of two condensates, 
where two phases, the weakly and strong 
segregated phases, characterized by the healing length and the penetration 
depth, have been predicted \cite{ac}.
We shall  use Eq. (1) and (2) under the condition (3) to study a highly 
non-linear dynamical process: 
The two condensates are initially in a homogeneously mixed state,
then eventually approach the phase segregated state.
In the same mean field treatment as in Ref.\onlinecite{ac}, 
we find that this dynamical
process can be classified into two main stages:
The initial highly non-equilibrium dynamical growth in the stage I, where the
dynamics is governed by the fastest growth mode, and the stage II of 
approach to equilibrium where the dynamics 
is governed by the slowest mode.
The stage II is typical of a relaxation process near equilibrium.
However, we shall show again it is governed by a quantum effect, namely,
the Josephson effect.

{\it Stage I:  Fastest Growth Mode. }
The coupled non-linear Schr\"odinger equations
have an obvious homogeneous solution:
Inside the trap the condensate densities $|\psi_j|^2 = \rho_{j0} $,
$
   \rho_{j0} =  {N_j }/{V} \; , 
$
with $V$ the volume of the square well potential trap,
and the chemical potentials
$
   \mu_1 = G_{11} \rho_{10} + G_{12}\rho_{20}   
$
and
$ 
   \mu_2 = G_{22} \rho_{20} + G_{12}\rho_{10}   
$.
This is the initial condition of the present problem.

To look for the fastest growth mode out of the homogeneous state,
we start with small fluctuations from the homogeneous state.
This is consistent with the usual stability analysis \cite{law}.
Our approach here is to emphasize the connection with the
physics of the classical spinodal decomposition and the role played by the 
Josephson relationships. Define
\be
   \psi_j(x,t) = \sqrt{\rho_j(x,t) } \; e^{i\theta_j(x,t) } \; ,
\ee 
and define the density fluctuations $\delta \rho_j = \rho_j - \rho_{j0} $
and the phase fluctuations  $ \theta_j $, and 
assume they are small: 
$  |\delta\rho_j | / \rho_j \; , \; |\theta_j | << 1 $.
The definition of the phase fluctuations here has made use of
the implicit assumption that there is no net current in the condensate.
To the linear order, we have from Eqs.(1,2), after eliminating 
the phase variables,
\be
   \frac{ \partial ^2 }{\partial t^2 } 
     \left( \begin{array}{c} 
            \delta\rho_1 \\
            \delta\rho_2  \end{array}   \right)
   = \left( \begin{array}{cc} 
       b_1 & \frac{ \rho_{10}}{m_1} G_{12 } \nabla^2  \\
                  \frac{ \rho_{20}}{m_2} G_{12 } \nabla^2 & b_2  
         \end{array}  \right) 
      \left( \begin{array}{c} 
            \delta\rho_1 \\
            \delta\rho_2  \end{array}   \right)  \; .
\ee
with 
$
  b_j = -\frac{\hbar^2}{4m_j^2}\nabla^4 + \frac{\rho_{j0}}{m_j} G_{jj}\nabla^2
$.
We look for the solution of the form
\[
   \left( \begin{array}{c} 
            \delta\rho_1 \\
            \delta\rho_2  \end{array}   \right)
   = \left( \begin{array}{c} 
            A \\
            B  \end{array} \right) e^{i ({\bf q}\cdot {\bf r} - \omega t)} \; ,
\]
with $A,B$ constants.
There are two branches of solution for Eq. (5).
For one branch, the frequency is always real. 
For another branch which we denote by $\omega_{-}$, 
it can be imaginary.
An imaginary frequency
$\omega_{-}$ shows that the initial homogeneously mixed state
is unstable.
It is straightforward to verify 
that the sufficient condition for the appearance of imaginary frequency 
for small enough wave number $q$ 
is the validity of the inequality Eq.(3).
The modes with imaginary frequencies will then
grow exponentially with time.
Unlike the usual situation near equilibrium, 
this growth from the present 
non-equilibrium homogeneously mixed state 
will be dominated by the fastest growth mode.
This is precisely the same case as in the initial stage of 
the classical spinodal decomposition.

To get a concrete understanding of the physical implications of 
Eq. (5) in the stage I 
we consider a case relevant to recent experiments where
particles of the two condensates have the same mass,
$ m_1 = m_2 = m$.
In this case we find the wavenumber corresponding to the {\bf most negative}
$\omega_{-max}^2$ is 
\be
   q_{max}^2 = \frac{m}{\hbar^2 } 
     \left[ \sqrt{ b_0^2
            + 4 {\rho_{10}\rho_{20} } G_{11}G_{22} 
                \left( \frac{G_{12}^2 }{G_{11}G_{22} } - 1 \right)  }
          -  b_0 \right] \; ,
\ee
and 
\be
   \omega_{-max} = - i \frac{\hbar }{2m} q_{max}^2   \; ,
\ee
with
$
 b_0 =  \rho_{10} G_{11} + \rho_{20} G_{22} 
$.
The physics implied in Eqs. (6,7) is as follows.
Starting from the initial homogeneous mixture of the two condensates, 
on the time scale given by 
\be
  t_I = 1/|\omega_{-max}| \; ,
\ee 
domain patterns of the phase 
segregation with the characteristic length 
\be
  l_I = 1/q_{max}
\ee
will appear.  Particularly, for the weakly segregated phase of 
$ {G_{12}^2 }/{G_{11}G_{22} } - 1 \rightarrow 0 $, we
have the length scale 
$
  l_I^{-1} = { \sqrt{2} }/{ \sqrt{ \Lambda_1^2 + \Lambda_2^2 } }  
$,
and for the strongly segregated phase  of
$ {G_{12}^2 }/{G_{11}G_{22} } - 1 \rightarrow \infty$,
we have 
$
  l_I^{-1} = {\sqrt{2}} / { \sqrt{ \Lambda_1 \Lambda_2 } } 
$.  
Here  $ \Lambda_j = \xi_j /\sqrt{ {G_{12} }/{\sqrt{ G_{11}G_{22} } } - 1 }$
and $\xi_j = \sqrt{ {\hbar^2}/{2m_j} \; {1}/{ \rho_{j0}G_{jj} } }$ 
are the penetration and healing lengths in the binary BEC mixture 
\cite{ac}. 
Those length and times scales can be measured experimentally.
We will come back to the experimental situation below.

After the stage I of fast growth into the domain 
pattern characterized by the length scale $l_I$, 
the system will gradually approach 
the true ground state of the complete phase segregation: 
one condensate in one region and the second condensate in another region.
This stage is slow, dominated by the slowest mode, and is 
the subject of stage II. 

It is now evident that the stage I of the growth of 
binary BEC's 
shares the same phenomenology of the initial stage of the classical
spinodal decomposition: the domination of the fastest growth mode, 
the appearance of domains of segregated phases, and
the conservation of particle numbers.
There are, however, two important differences.
First, the dynamical evolution of the binary BEC's is governed by
a coupled nonlinear time dependent Schr\"odinger equations,
not by a nonlinear diffusion equation supplemented with 
the continuity equation, the Cahn-Hilliard equation.
There is no external relaxation process  for the present wave functions.
Secondly, 
the energy of the binary BEC's is conserved during the growth process,
not as in the case of classical spinodal decomposition where
the system energy always decreases.
  
{\it Stage II: Merging and Oscillating between Domains. }
The BEC binary mixture occurs in a trap.
This finite size effect of the droplet
leads to the broken symmetry of the condensate
profiles \cite{cta}, 
which tends to separate the condensates in mutually isolated regions.
This implies that 
there is no contact between different domains of the same
condensates formed in the stage I.
The classical spinodal process involves diffusion. An estimate of the
diffusion constant for the BEC system can be made from kinetics theory.
The ratio of the time scales for quantum and classical particle
transport is of the order of the ratio of the BEC cloud size
to the de Broglie wavelength.
This is much smaller than one for the experimental systems of interest.
The classical diffusion process is thus not important.
Because the domains are not connected 
and because the diffusion for the BEC mixture is
extremely low, all the mechanisms for the late stage classical
spinodal decomposition are not applicable. 
We propose that it is the Josephson effect that is responsible for the
approach to equilibrium in the stage II.
Two models for the Josephson effect, the `rigid pendulum' model and 
the `soft pendulum' model will be discussed.
They both give the same time scale when the `Rabi' frequency is small.

Let us consider the specific case of two domains of 
 condensate 1 separated by a domain with width $d$ of  condensate 2.
The ability of condensate 1 to tunnel through condensate 2
is described by the penetration depth $\Lambda$, 
as discussed in Ref. \onlinecite{ac}.
Hence the probability of condensate 1 to tunnel through condensate 2 
can be estimated as
\be
   p =   e^{ - 2  {d}/{\Lambda} }  \; ,
\ee
when $p$ is much smaller than 1.
The finiteness, though small, of the tunneling probability suggests 
that it is the Josephson effect responsible for the relaxation process in
the stage II. 
The Josephson effect 
leads to the merging of two domains of the same condensate.
The dynamics of such motion may be governed by the `rigid pendulum' 
Hamiltonian for a Josephson junction \cite{tunneling1}:
\[
   H(\phi, n ) = E_J ( 1- \cos \phi ) + \frac{1}{2} E_C \;  n^2 \; ,
\]      
where $E_J$ is the Josephson coupling energy
determined by the tunneling probability,
$n= (n_x - n_y )/2$ is the particle number difference between 
the numbers of particles, $n_x$ and $n_y$, in the two domains,
and $E_C \equiv \partial \mu /\partial n $  is the `capacitive' energy
due to interactions. 
In the absence of external constraints, $ \mu = E_C n $.
The phase difference $\phi$ between the two domains is conjugated to $n$,
as in usual Josephson junctions.
Under the appropriate condition, such as low temperature and smallness of the
capacitive energy,  
there may be an oscillation  between the two domains of  condensate 1
separated by condensate 2.
In such a case, we may estimate that 
the oscillation period $ t_{II} = 2\pi/\omega_{JP} $, with the so-called
Josephson plasma frequency\cite{tunneling1}
\be
   \omega_{JP} =  \frac{ \sqrt{ E_C E_J }}{\hbar} \; .
\ee
For small tunneling probability, the Josephson junction energy may be
estimated as\cite{tunneling1}
$
   E_J = n^{1/3}_{T} \hbar\omega_0 \;  e^{ - 2 {d}/{\Lambda} } 
$,
and the capacitive energy as
$
   E_C = \frac{2}{5} \left( {n_T }/{2} \right)^{-0.6} 
          \left( {15a_{11} }/{a_{0} }\right)^{0.4} {\hbar\omega_{0}} 
$.
Here $\omega_0$ is the harmonic oscillator frequency for  condensate 1
     in a harmonic trap, 
     $a_0 =  \sqrt{\hbar/m_1\omega_0} $ is the corresponding oscillator length,
     the $a_{11}$ is the scattering length of  condensate 1, and 
     $n_T = n_x + n_y$ is the total number of particles in domain $x$ and $y$.
The oscillatory time scale between the domains determines by 
the Josephson plasma frequency 
\be
   t_{II}^{-1} = \left( \frac{2a_0 }{15a_{11} \; n_T } \right)^{2/15} 
      \frac{\omega_0}{2\pi}  \;  e^{ - {d}/{\Lambda} }\; .
\ee
The rigid pendulum Hamiltonian would give a good description when
$n << n_T$.
Another description of the Josephson effect uses the 
`soft pendulum' Hamiltonian proposed in Ref. \onlinecite{tunneling2}.
In the parameter regime relevant current experimental situations both of 
them give the same order of magnitude estimate for $t_{II}$.

Given the Josephson effect is the dominant
mechanism in the stage II, 
the time scale  to arrive at the ground state
will be determined by the Josephson effect at the final two domains, 
in which the two domains of condensate 1 is separated by 
the domain of  condensate 2, in the manner of $1 \, 2 \, 1 \, 2 $
spatial configuration 
for the case of equal numbers of the condensates.
The width $d$ of each domain is then $D/4$, with $D$ the size of the trap.  
According to the above analysis, 
the largest time scale determined by Eq.(12), the slowest mode, 
in the stage II is $t_{II}$.
The arrangement of $1\, 2 \, 1 $ spatial configuration 
may also occur here, in which it is
more likely for  condensate 2 to tunnel through 1 to the edge of the trap,
because of the larger tunneling probability. We next turn our attention
to the experimental situation.

The first question is whether the quantum spinodal decomposition 
discussed above can happen or not.
In terms of the atomic scattering lengths of condensate atoms
$a_{jj}$, the interactions are 
$G_{jj} = {4\pi \hbar^2 a_{jj} }/{m_j } $.
The typical value of $a_{jj}$ for $^{87}$Rb is 
about $50$\AA.
The typical density realized for the binary BEC mixture is 
about $\rho_{i0} \sim 10^{14}/cm^3$.
Hence the healing length is 
$\xi = \sqrt{  ({\hbar^2}/{2m}) ({1}/{G_{jj} \rho_{j0} } ) }
     = \sqrt{1/(8\pi a_{jj}\rho_{j0} ) } \sim 3000$\AA.
For the different hyperfine states of $^{87}$Rb,
it is known now\cite{jila2,cta} that 
${G_{12} }/{ \sqrt{ G_{11}G_{22} } } > 1$.
Hence the ground state of the phase segregated phase can be realized.
Also, that both the healing length and the penetration depth are smaller than
the trap size \cite{ac} validates the application of results obtained 
in the square well trapping potential to the harmonic one.
Therefore, the quantum spinodal decomposition can happen.

Experimentally, starting from the initially homogeneous state,
after a short period of time 
a domain pattern does appear.
Then a damped oscillation between the domain pattern has been observed.
Eventually the binary BEC mixture sets to the segregated phase \cite{jila2}.
If we take ${G_{12} }/{ \sqrt{ G_{11}G_{22} } } = 1.04$,
 the penetration depth is 
$\Lambda = \xi/\sqrt{ {G_{12} }/{ \sqrt{ G_{11}G_{22} } } - 1 }  
     \sim 1.5 \mu$m.
The length scale $l_I$ determined by Eq.(9) in the stage I 
is $ 1.5 \mu$m, which is the same order of magnitude in comparison with
experimental data\cite{jila2}.
This length is also comparable to the domain wall width seen
experimentally.
The corresponding time scale $t_I$ (Eq. (8)) is then 6 ms, again the 
same order of magnitude, 
though both estimated $l_I$ and $t_I$ are somewhat smaller.
For stage
II, $d/\Lambda$ is of the order of $n_d$, the number of domains formed
in stage I. For the experiment of interest to use, $n_d\approx 2$.
If we assume that the damped oscillation to equilibrium observed experimentally
is the stage II discussed here, 
taking the total number of particle $N_1 = 10^{6}$,
we find the period according to Eq. (12) is 30 ms, comparable to the 
experimental value.

For the stage I, our above analysis shows that 
it is insensitive to the damping. 
At this moment we do not have a reliable estimation of the damping
in the Josephson oscillation, 
whose existence has been indicated experimentally.
Nevertheless,
given the uncertainty in the value of $G_{12} $ or $a_{12}$,
we conclude that the stage II of the quantum spinodal decomposition
may have been observed.  
 

Periodic-like structures have also been observed in the phase
segregation of spin 1 Na mixtures \cite{mit}. 
For the definiteness of analysis we have presently focused on the binary
case, but we believe a similar
physical picture of quantum spinodal decomposition 
applies to that case as well.

Finally, we point out that the length and
time scales for Rb mixtures is controlled by the difference of
$r=G_{12}/\sqrt{G_{11}G_{22}}$ and 1. Since $r$ is close to 1
experimentally, the data can provide for a very sensitive estimate of
r-1. Since this parameter determines $\bf both$ the time and length
scale of stage I, it is a self-consistent check on the physical picture
provided here.

{\ }

\noindent
{ This work was supported in part  by a grant from NASA (NAG8-1427), and 
  by Swedish NFR (PA). 
  One of us (PA) thanks the Bartol Research Institute as well 
  as the Department of Physics at University of Delaware for the pleasant 
  hospitality, where the main body of the work was completed.
  We also thank the D. S. Hall for sending us their data. }

\end{document}